# MATLAB-VISSIM INTERFACE FOR ONLINE OPTIMIZATION OF GREEN TIME SPLITS

**Prateek Bansal**
**Graduate Research Assistant**
**Department of Civil, Architectural and Environmental Engineering**
**The University of Texas at Austin**
**Email:** prateekbansal@utexas.edu

## ABSTRACT

VISSIM is a widely used microscopic traffic simulator, which not only provides a graphical user interface to simulate simple static controls (pre-timed or fixed-time) but also offers flexibility to dynamically control simulation through versatile programming languages (C++ and Java etc.). However, to implement various traffic control techniques, integration of computational tools may save lots of effort and time as compared to standard programming platforms. MATLAB falls in the category of a widely used computational tool and also fulfills the primary requirements to control VISSIM simulation dynamically. Therefore, this study proposes and develops a **direct** interface between MATLAB and VISSIM to extensively harness the computational power of MATLAB. The significance of developed interface is demonstrated on a practical scenario, by conducting an online optimization of green time splits on a study network.

**SCOPE AND OBJECTIVES**

To manage a rapid growth in travel demand, various traffic control algorithms (artificial neural networks and genetic reinforcement learning etc.) and management strategies evolved in recent years, but validation of these algorithms/strategies is still a key challenge. Field experiments are most effective and direct way to evaluate these algorithms/strategies but they demand for substantial time, money and effort. In the current paradigm of information technology, microscopic traffic simulation seems to be the most efficient and fastest way to validate algorithms/models. VISSIM provides a microscopic traffic flow simulation environment to analyze traffic dynamics by replicating real world traffic conditions (lane settings, traffic composition, traffic signals, bus stations, and driving behavior etc.). The simulator uses theoretical psycho-physical driver behavior model developed originally by Wiedemann (1). Though, VISSIM provides a graphical user interface (GUI) to simulate simple static controls (fixed-time), but it has various limitations. User can't access and manipulate ongoing simulation through GUI (2). Therefore, VISSIM offers a COM interface having hierarchical structure of objects, properties, and methods, which can be manipulated dynamically by versatile programming languages (C++ and Java etc.) having ability to handle COM objects (3).

To implement various traffic control techniques in C++ or Java, rigorous programming efforts and time investment are essentially required. Moreover, in several cases, for e.g. optimization procedures, users may not be required to know the entire mathematical intricacies of an algorithm and thus, an integration of a well-established optimization tool may save lots of energy and time as compared to programming in any language. Among such computational tools, MATLAB is widely used mathematical software and includes a large number of computation toolboxes and library functions, which allow easy implementation of a variety of complex control algorithms without knowing their minor details (4). MATLAB also fulfills the primary requirements to develop an interface with VISSIM by having the ability to handle COM objects, which would allow its online communication with VISSIM during simulation.

The interaction between VISSIM and MATLAB has not been extensively explored in the literature. As per author's best knowledge, only one recent study (4) explored application of interaction between VISSIM and MATLAB, in which MATLAB routines were called through VISSIM COM interface. In contrast to the earlier study (4), this study considers MATLAB as a client, VISSIM as a server and there is a client server interaction i.e. MATLAB controls microscopic simulation rather than being controlled by the COM interface. To harness the computational power of MATLAB extensively, this study develops a **direct** interface between MATLAB and VISSIM. For illustrating advantage of MATLAB-VISSIM interface, a traffic-responsive urban control (TUC) strategy has been implemented to reduce total system delay by online optimization of green time splits among different phases of traffic controllers in the network.

**TUC STRATEGY FOR ONLINE OPTIMIZATION OF GREEN TIME SPLITS**

This study uses the split control part of the TUC signal control strategy, derived from a formulation in the format of a linear-quadratic (LQ) control problem, leading to the multivariable regulator (5) as shown in Equation 1.

$$g(k) = g^N - Lx(k) \tag{1}$$

Where, $g^N$ is a vector of nominal green time corresponding to a pre-specified fixed signal plan, $x(k)$ is a vector of vehicle-numbers on links approaching intersection at the start of cycle k, i.e. at the end of the previous cycle k −1, and **L** is a constant gain matrix. This study uses Equation 1 to calculate new optimized green time splits after completion of each signal cycle. L is most crucial variable of Equation 1 and calculated offline using inbuilt "LQ state-feedback regulator for discrete-time state-space system" in MATLAB as shown in Equation 2.

$$[L, \sim, \sim] = dlqr(A, B, Q, R) \tag{2}$$

The study network (Figure 1) has 2 signal controllers, each having 3 phases i.e. 6 control variables and 6 links approaching to the intersections i.e. 6 state variables. Therefore, Matrix A = 6x6 identity matrix; Matrix R = r*(6x6 identity matrix) where r=.0001 to .000001; Matrix Q = 6x6 diagonal matrix where diagonal elements are inverse of storage capacity of corresponding links approaching to intersection. To find matrix B, a mathematical model of TUC strategy (6) was formulated for the network used in this study. The complicated state equation has been simplified to Equation 3, where Δg(k) is a vector of the difference between pre-specified green time and green time at time step k. Scalar form of Equation 3 has been written for all 6 links approaching to intersection and later, combined to extract **B** matrix.

$$x(k+1) = x(k) + B\Delta g(k) \tag{3}$$

It is important to note that LQ regulator doesn't take into account for two inherent constraints- a) constant cycle length, and b) lower & upper bound on effective green time. The green time obtained from Equation 1 is modified appropriately to account for these constraints by solving a quadratic programming problem (QPP) in real time (after each cycle) (5). An inbuilt function (quadprog) of MATLAB optimization toolbox has been used to solve the QPP.

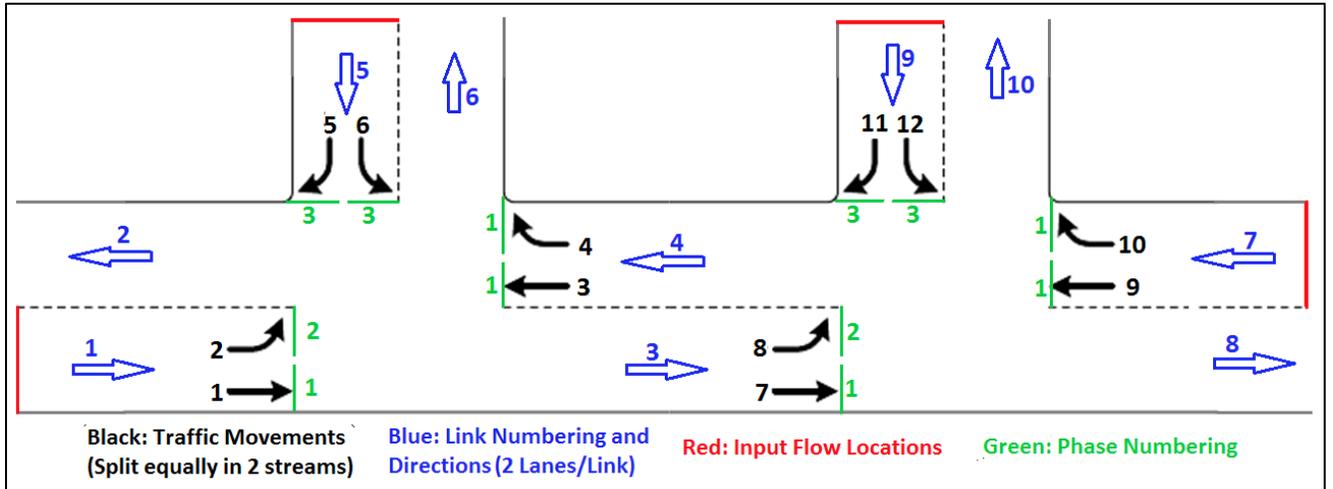

**Figure 1** Specifications of the network employed in the study

# MATLAB-VISSIM INTERFACE (ONLINE OPTIMIZATION OF GREEN TIME SPLITS)

A flow-chart (Figure 2) depicts the entire process of developing direct MATLAB-VISSIM interface for online optimization of green time splits. Irrespective of using any programming language or MATLAB to control VISSIM simulation dynamically, creation of traffic network and corresponding settings have to be done through VISSIM GUI. Network and corresponding settings are saved in VISSIM project (INP) and initialization file (INI) respectively. After creation of these two files, entire VISSIM microscopic simulation is controlled dynamically by MATLAB environment. To develop direct MATLAB-VISSIM interface, two MATLAB programs were written- a) to control VISSIM simulation dynamically (Sim-Prog) and, b) to optimize green time splits after completion of each cycle (Green-Opt).

In Sim-Prog, VISSIM server is first activated to realize COM object and then, INP & INI files are taken as input. Then, 'network' and 'simulation', two top hierarchical objects, are accessed through COM object. Later, network settings can be realized and modified in MATLAB environment by accessing methods and properties (length of link and state of signal etc.) of lower order hierarchical objects (link and signal controller etc.) through a 'network' object. For this study, a single step microscopic simulation with step size of 1 second has been chosen to control and set the state of signal controllers dynamically. After completion of each cycle (cycle length=60 seconds), Green-Opt is called from Sim-Prog. A vector of number of vehicles on the links approaching intersection ($x(k)$), constant gain matrix (L), and pre-specified green time vector ($g^N$) are passed as input to Green-Opt, and new optimized green time vector for next cycle ($g(k)$) is obtained as output. The states of signals in next cycle are set according to the optimized green time vector and the entire process continues till the end of total simulation time, which was 3000 seconds (50 cycles) for this study. In the end, total system delay is reported (Table 1) by adding cycle delays of all routes in the network (accessed by 'delay' object).

To check the effectiveness of green time splits optimization algorithm, microscopic simulation has been executed for fixed time signal scenario under same network settings and resultant delays are reported in Table 1. For microscopic simulation in fixed time signal scenario using MATLAB-VISSIM interface, similar methodology (Figure 2) has been followed by escaping "calling of Green-opt function" from Sim-Prog after each cycle. Though it is not a central focus of the study, but significant differences in system delays (Table 1) are visible for the 'fixed time' and 'green time optimized' scenarios, reflecting success of TUC strategy implementation using developed interface.

**Table 1** Summarizing result of traffic responsive and fixed time strategy

| Sr. No. | Vehicle input (vehicles/hour) | | | | Total system delay (in seconds) | |
|---|---|---|---|---|---|---|
| | Link 1 | Link 5 | Link 9 | Link 7 | Green time optimized | Fixed time |
| 1 | 1000 | 1000 | 1000 | 1000 | 780 | 1145 |
| 2 | 1200 | 1000 | 1000 | 1200 | 1670 | 2060 |
| 3 | 1200 | 1200 | 1200 | 1200 | 1650 | 2277 |

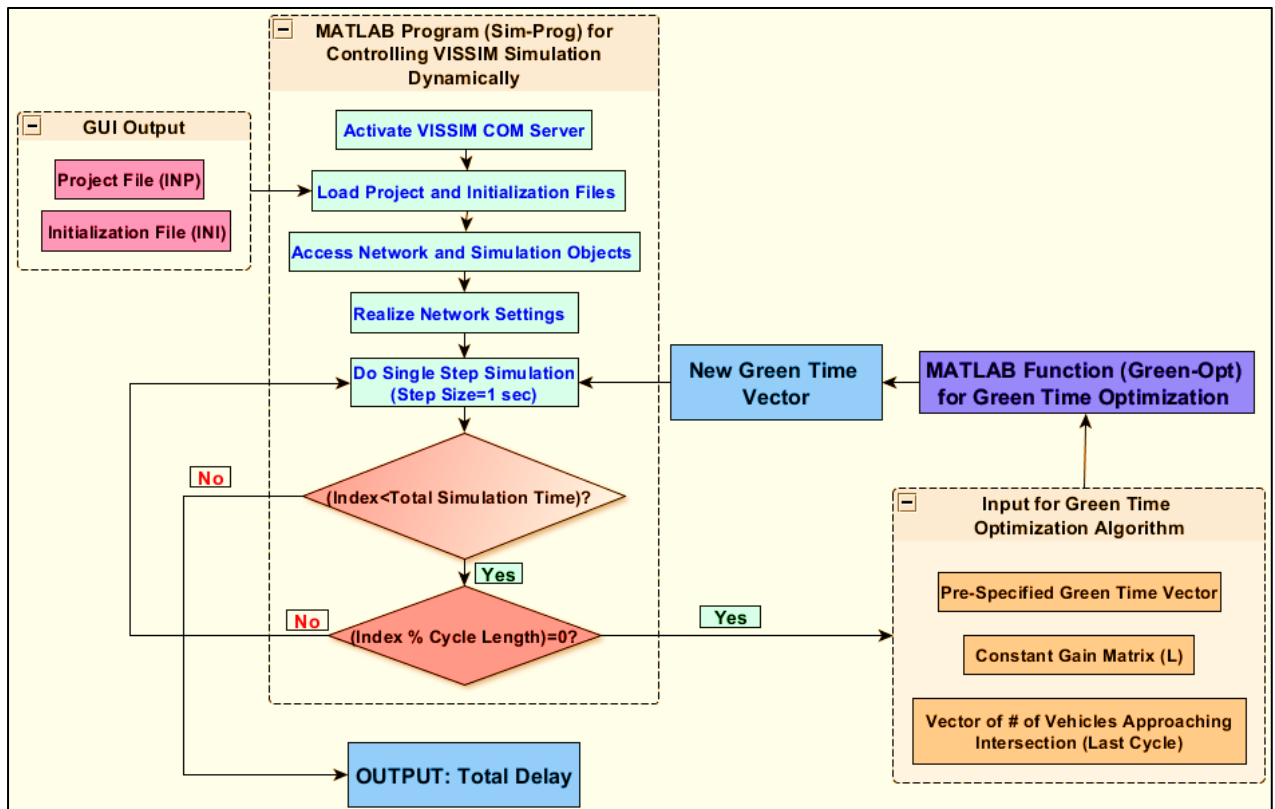

**Figure 2** MATLAB-VISSIM interface for green time optimization

## CONCLUSIONS AND FUTURE WORK

The study proposed a methodology to dynamically control microscopic simulation in VISSIM through MATLAB and demonstrated it on a study network for online optimization of green time splits. The proposed interface may help in building quick simulation platform for various control strategies by reducing a lot of programming effort. By integrating advanced MATLAB toolboxes during simulation, several researchers would be benefited as they need not understand intricacies of the algorithms and can instead channelize their efforts on the modeling and practical aspects .

It is important to note that a real time implementation of the demonstrated TUC strategy using MATLAB-VISSIM interface needs parallel processing of Sim-Prog and Green-Opt because there is an online exchange of data between both programs during simulation (after each cycle). From simulation stand point, parallel processing is not a requirement because VISSIM simulation stops during processing of optimization function (Green-Opt) due to pause of MATLAB program (Sim-Prog) controlling microscopic simulation. For real time implementation of proposed approach, various parallel computing tools can be explored. MATLAB Message Passing Interface (MPI) and parallel computing toolbox come under the set of such available options.

## ACKNOWLEDGEMENTS

This work has been funded by Dr. Balazs Adam Kulcsar, Assistant Professor, Department of Signals and Systems, Chalmers University of Technology, Gothenburg, Sweden. I thank to Dr. Kulcsar and also owe thanks to Ms. Azita Dabiri (Doctoral Researcher) for guiding me.

## REFERENCES

(1) Wiedemann, R. "Simulation des Straßenverkehrsflusses, Schriftenreihe Heft 8."*Institute for Transportation Science, University of Karlsruhe, Germany* (1994).

(2) PTV, AG. "Vissim 5.40-01 user manual." *Karlsruhe, Germany* (2011).

(3) Roca, V. "User Manual for the VISSIM COM interface." *PTV Planung Transport Verkehr AG* (2005).

(4) Tettamanti, Tamds, and Istvdn Varga. "Development of road traffic control by using integrated VISSIM-MATLAB simulation environment." *Periodica Polytechnica Civil Engineering* 56*, no. 1* (2012): 43-49.

(5) Kouvelas, Anastasios, Konstantinos Aboudolas, Markos Papageorgiou, and Elias B. Kosmatopoulos." A hybrid strategy for real-time traffic signal control of urban road networks." *Intelligent Transportation Systems, IEEE Transactions on 12*, no. 3 (2011): 884-894.

(6) Diakaki, Christina, Markos Papageorgiou, and Kostas Aboudolas. "A multivariable regulator approach to traffic-responsive network-wide signal control." *Control Engineering Practice 10, no. 2* (2002): 183-195.

---

If you want to access MATLAB programs (Sim-Prog & Green-opt), Please email me at prateekbansal@utexas.edu.